\documentclass[superscriptaddress,aps,journal=prl,preprint]{revtex4-2}
\usepackage{graphicx}
\usepackage{epstopdf}
\usepackage{bm,amsmath,amssymb} 
\usepackage{color, soul} 
\usepackage{dcolumn}   
\usepackage{multirow}
\usepackage{makecell}
\usepackage{threeparttable}

\newcommand{\eg}{{\em e.\,g.}}

\newcommand{\etal}{{\em et al.\ }}
\newcommand{\comment}[1]{}

\begin{document}
\title{Onsite and intersite electronic correlations in the Hubbard model for halide perovskites}

\author{Jiyuan Yang}
\thanks{These two authors contributed equally}
\affiliation{Zhejiang University, Hangzhou, Zhejiang 310058, China}
\affiliation{Key Laboratory for Quantum Materials of Zhejiang Province, Department of Physics, School of Science, Westlake University, Hangzhou, Zhejiang 310024, China}
\author{Tianyuan Zhu}
\thanks{These two authors contributed equally}
\affiliation{Key Laboratory for Quantum Materials of Zhejiang Province, Department of Physics, School of Science, Westlake University, Hangzhou, Zhejiang 310024, China}
\affiliation{Institute of Natural Sciences, Westlake Institute for Advanced Study, Hangzhou, Zhejiang 310024, China}
\author{Shi Liu}
\email{liushi@westlake.edu.cn}
\affiliation{Key Laboratory for Quantum Materials of Zhejiang Province, Department of Physics, School of Science, Westlake University, Hangzhou, Zhejiang 310024, China}
\affiliation{Institute of Natural Sciences, Westlake Institute for Advanced Study, Hangzhou, Zhejiang 310024, China}


\begin{abstract}{ 
Halide perovskites (HPs) are widely viewed as promising photovoltaic and light-emitting materials for their suitable band gaps in the visible spectrum. Density functional theory (DFT) calculations employing (semi)local exchange-correlation functionals usually underestimate the band gaps for these systems. Accurate descriptions of the electronic structures of HPs often demand higher-order levels of theory such as the Heyd-Scuseria-Ernzerhof (HSE) hybrid density functional and $GW$ approximations that are much more computationally expensive than standard DFT. Here, we investigate three representative types of HPs, $ABX_3$ halide perovskites, vacancy-ordered double perovskites (VODPs), and bond disproportionated halide perovskites (BDHPs), using DFT+$U$+$V$ with onsite $U$ and intersite $V$ Hubbard parameters computed self-consistently without {\em a priori} assumption. 
The inclusion of Hubbard corrections improves the band gap prediction accuracy for all three types of HPs to a similar level of advanced methods. Moreover, the self-consistent Hubbard $U$ is a meaningful indicator of the true local charge state of multivalence metal atoms in HPs. The inclusion of the intersite Hubbard $V$ is crucial to properly capture the hybridization between valence electrons on neighboring atoms in BDHPs that have breathing-mode distortions of halide octahedra. In particular, the simultaneous convergence of both Hubbard parameters and crystal geometry enables a band gap prediction accuracy superior to HSE for BDHPs but at a fraction of the cost. 
Our work highlights the importance of using self-consistent Huabbard parameters when dealing with HPs that often possess intricate competitions between onsite localization and intersite hybridization.}
\end{abstract}

\maketitle
\newpage

\section{Introduction}

Halide perovskites (HPs) have drawn intensive interest over the past decade for their remarkable optoelectronic properties and promising photovoltaic applications~\cite{Kojima09p6050,Lee12p643,Jeon15p476,Green22p3}. The suitable and tunable band gap and small charge carrier effective mass of HPs make them excellent candidates for optical absorber materials~\cite{Noh13p1764,Yin14p4653}. 
Since the report of the very first perovskite-based solar cell (PSC) by \citeauthor{Kojima09p6050} in 2009, the power conversion efficiencies (PCEs) of PSCs have improved rapidly from 3.8\% to 25.5\%~\cite{Kojima09p6050,Green22p3}. Despite the high performance and the low fabrication cost of HPs, the intrinsic long-term instability remains a major hurdle impeding the commercialization of PSCs~\cite{Correa-Baena17p739,Saliba18p388}. In addition, the presence of toxic elements such as lead (Pb) raises concerns regarding the environmental and health impacts of large-scale deployment of Pb-based PSCs~\cite{Im19p37,Wu21p863}. All these issues have prompted the scientific community to search for different types of perovskite derivatives~\cite{Xiao19p1803792}.

\begin{figure}[]
\centering
\includegraphics[width=1\textwidth]{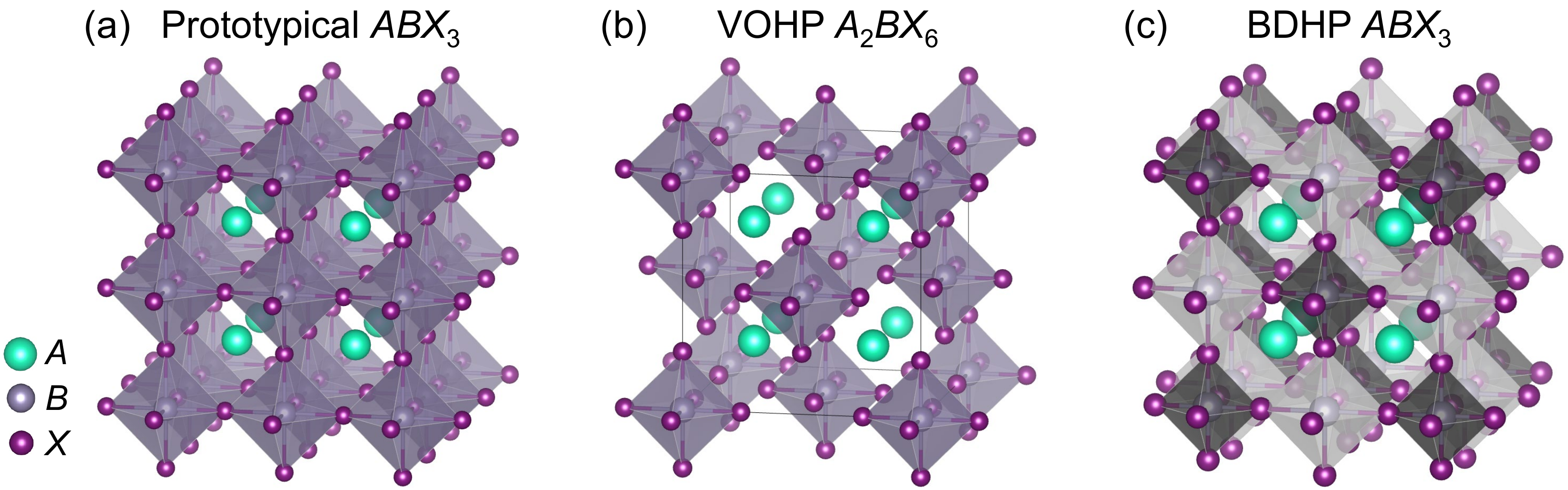}
\caption{Schematics of three types of halide perovskites. (a) Prototypical $ABX_3$ HP, (b) $A_2BX_6$ VODP, and (c) $ABX_3$ BDHP.}
\label{struc}
\end{figure}

The prototypical HP  has a general chemical formula of $ABX_3$, where $A$ is a monovalent organic or inorganic cation (\eg, CH$_3$NH$_3^{+}$ and Cs$^{+}$), $B$ is a divalent metal cation (\eg, Pb$^{2+}$ and Sn$^{2+}$), and $X$ is a halide anion (\eg, I$^{-}$, Br$^{-}$, and Cl$^{-}$). The $ABX_3$ structure consists of a corner-sharing $BX_6$ octahedra network with the $A$-site cations locating in the cavities between adjacent octahedra (Fig.~\ref{struc}a). To address the long-term instability issue of CsSnI$_3$ in which Sn adopts a formal oxidation state of 2+, \citeauthor{Lee14p15379}~introduced Cs$_2$SnI$_6$, a vacancy-ordered double halide perovskite (VOHP), that has Sn adopting the 4+ formal oxidation state and is stable in air and moisture~\cite{Lee14p15379}. The VOHP is derived from the $ABX_3$ HPs by removing every other $B$-site cations (Fig.~\ref{struc}b), and hence have a chemical formula of $A_2BX_6$ \cite{Xiao15p1250,Maughan19p1184}. By forming vacancies at half of the $B$ sites, the nominal oxidation state of $B$ changes from $2+$ in $ABX_3$ to $4+$ in $A_2BX_6$. Since then, various types of VOHPs such as Cs$_2BX_6$($B=$ Sn, Te, Ti, Zr, Pd, Pt) in which the $B$ site is occupied by metals with stable oxidation state of $4+$ have been synthesized \cite{Kaltzoglou16p11777,Maughan16p8453,Folgueras21p25126,Ju18p297,Kong20p1591,Abfalterer20p1644,Zhou18p2613,Sakai17p6030,Wang20p13310,Schwartz20p2000182}.

Some HPs such as CsAuCl$_3$~\cite{Matsushita07p1353} and CsTlF$_3$~\cite{Retuerto13p4071} possess so-called breathing-mode 
distortions of halide octahedra where $BX_6$ cages are alternately dilated and contracted in a rocksalt-like pattern (Fig.~\ref{struc}c). The breathing  structural distortions were originally attributed to the charge disproportionation of $B$-site cations, $2B^{2+}\rightarrow B^{+}+B^{3+}$, that leads to two different sets of $B$--$X$ bond lengths~\cite{Varma88p2713,Hase07p174103,Liu99p7925}. This type of structural flexibility enriches the optical and electronic properties of halide perovskites such as high-$T_C$ superconductivity and semiconductor-to-metal phase transition \cite{Liu99p7925,Retuerto13p4071,Wang13p054104,Lin19p23404}.
Recently, such charge-ordering picture was challenged both theoretically and experimentally~\cite{Raebiger08p763,Dalpian18p075135, Benam21p195141}. For example, investigations based on density functional theory (DFT) calculations indicated that the physical charges of $B$-site cations in CsAuCl$_3$ remain largely unchanged despite being in different local environments~\cite{Dalpian18p075135, Benam21p195141}. This is due to the ``charge self-regulating response"~\cite{Raebiger08p763} that $B$-site metal atoms and $X$-ligands rehybridize to prevent the creation of highly charged cations through the ligand-to-metal charge transfer. For this reason, it was proposed to describe these $ABX_3$ HPs as ``bond disproportionated" (BD) instead of ``charge ordered". 
Similarly, it was found that the actual charge density residing on the Sn atom in CsSnI$_3$ and Cs$_2$SnI$_6$ is nearly identical although Sn adopts different formal oxidation states in these two compounds~\cite{Dalpian17p025401}. 

An accurate description of the electronic structures of different types of HPs is vital for the rational design and optimization of HP-based photovoltaic and optoelectronic devices. DFT has played an important role in the discovery and design of HPs ~\cite{Even13p2999,Mosconi13p13902,Yin14p4653,Das22p2184}. However, the accuracy of DFT is limited by the approximations to the exchange-correlation energy functional. The remnant self-interaction error (SIE)~\cite{MoriSnchez06p201102} within the local-density approximation (LDA)~\cite{Perdew92p13244} and generalized gradient approximation (GGA)~\cite{Perdew96p3865} often causes a considerable underestimation of band gaps in systems with localized electronic states (\eg, 3$d$ electrons). It is noted that when the $B$-site is occupied by heavy elements such as Pb, the inclusion of the spin-orbit coupling (SOC) effects in DFT calculations can significantly reduce the theoretical band gap values~\cite{Even13p2999}.
As a result, there exists a notable error cancellation when using GGA without SOC to evaluate the band gaps of Pb-based HPs~\cite{Mosconi13p13902, Das22p2184}. Advanced methods such as $GW$ approximation and the Heyd-Scuseria-Ernzerhof (HSE) hybrid density functional are more accurate, but their expensive computational costs often limit the applications to small systems of tens of atoms. 

The DFT$+U$ method based on the mean-field Hubbard model that removes SIE for states in the Hubbard manifold is a cost-effective way to improve the descriptions of electronic correlations~\cite{Kulik15p240901}. The Hubbard parameter $U$ gauges the strength of the onsite (screened) Coulomb repulsion of localized electrons and is critical for the accuracy of DFT$+U$. In many studies, the value of $U$ is assumed to be element-specific and is manually tuned to reproduce experimental results and/or results from higher-order levels of theory. It is highly desirable to determine $U$ in an unbiased way with minimum human intervention. In order to address this problem, \citeauthor{Agapito15p011006} formulated a pseudohybrid Hubbard density functional, Agapito-Curtarolo-Buongiorno Nardelli (ACBN0), enabling a direct self-consistent computation of Hubbard $U$ parameters~\cite{Agapito15p011006}. By further including the Coulomb interactions between electrons on nearest-neighboring Hubbard sites (denoted as intersite $V$), Lee~\textit{et al.} and Tancogne-Dejean~\textit{et al.} respectively developed an extended version of ACBN0, termed as eACBN0, that enables a fully {\em ab initio} DFT+$U$+$V$ method using self-consistent onsite $U$ and intersite $V$ parameters~\cite{Lee20p043410,Tancogne-Dejean20p155117}.

It is desirable to develop a fully {\em ab initio} and computationally affordable approach to accurately predict the electronic structures of HPs. In this work, by employing ACBN0 and eACBN0, we perform an extensive investigation on the electronic structures of three different types of HPs: prototypical $ABX_3$ HPs, $A_2BX_6$ VODPs, and $ABX_3$ BDHPs, as sketched in Fig.~\ref{struc}. We compare the band gap values with those obtained in experiments and calculated by HSE and $GW$ methods (if available), and find that the self-consistent Hubbard $U$ and $V$ parameters affect the electronic structures of HPs to different degrees depending on the chemical bonding nature. For prototypical $ABX_3$ HPs, the DFT$+U$ method with $U$ computed with ACBN0 is sufficiently accurate, yielding band gap values at the $GW$ level; the intersite $V$, however, has limited impacts on the band structures. As regards VODPs, both $U$ and $V$ play significant roles, with DFT$+U$ correcting the band gap values to the HSE level and DFT+$U$+$V$ further to the $G_0W_0$ level. In the case of BDHPs,  we discover canceling effects when applying $U$ corrections to different localized states and the inclusion of the intersite Hubbard $V$ is crucial to capture the hybridization between valence electrons on neighboring Hubbard sites. Finally, we demonstrate that the simultaneous convergence of both Hubbard parameters and crystal geometry enables a band
gap prediction accuracy superior to HSE for BDHPs.

\section{COMPUTATIONAL METHODS}
We perform standard DFT calculations with \texttt{Quantum Espresso} (\texttt{QE})~\cite{Giannozzi09p395502,Giannozzi17p465901} using GGA of the Perdew-Burke-Ernzerhof (PBE) parametrization~\cite{Perdew96p3865}. Ultrasoft pseudopotentials taken from the Garrity-Bennett-Rabe-Vanderbilt (GBRV) library~\cite{Garrity14p446} are used for structural optimizations. For all inorganic HPs, we fully optimize both the lattice constants and atomic positions. In the case of organic-inorganic hybrid HPs such as CH$_3$NH$_3$PbI$_3$, we fix the lattice parameters to experimental values and fully relax the atomic positions. We use an energy convergence threshold of $10^{-7}$~Ry, a force convergence threshold of $10^{-6}$~Ry/Bohr, and a plane wave cutoff energy of 50 Ry to obtain the ground-state structures. An $8 \times 8 \times 8$ $k$-point mesh is used for cubic HPs, and the $k$-point mesh is modified accordingly for other phases to maintain a similar $k$-point density. For VODPs and BDHPs, a $4 \times 4 \times 4$ $k$-point mesh is used. 
Then based on PBE optimized structures, we compute $U$ and $V$ parameters with ACBN0 and eACBN0 using an in-house version of \texttt{QE}~\cite{Lee20p043410} and GBRV pseudopotentials. The threshold for the convergences of $U$ and $V$ values is set as $10^{-6}$~Ry. To include the SOC effects on the band gaps, fully relativistic optimized norm-conserving pseudopotentials provided by the PseudoDoJo project~\cite{Van18p39} are chosen, and an increased plane wave cutoff energy of 80 Ry is used to compute the band gaps. 
Because current implementation of DFT+$U$+$V$ in \texttt{QE} does not support fully relativistic pseudopotentials, the SOC-induced band gap correction in eACBN0+SOC is approximated as $\Delta_{\rm SOC}$ = $E_g$(DFT$+U+$SOC)$-E_g$(DFT$+U)$, where the values of $U$ in DFT+$U$ are computed self-consistently with ACBN0. That is, for a given HP, we assume the magnitudes of $\Delta_{\rm SOC}$ in ACBN0+SOC and eACBN0+SOC are the same.
For DFT+$U$ and DFT+$U$+$V$ calculations, we employ the simplified rotationally invariant formulation~\cite{Dudarev98p1505}, and the atomic orbitals orthogonalized using L{\"o}wdin's method~\cite{Timrov20p235159,Lowdin50p365} are adopted to construct projectors of the Hubbard manifold. The same settings of DFT+$U$+$V$ are used in the sc-eACBN0 loop (see later discussions) to converge both geometry and Hubbard parameters at the same time. The sc-eACBN0 calculation reaches the convergence when the changes in Hubbard parameters are less than $10^{-4}$~Ry. All structural files of optimized structures and representative input files for DFT+$U$ and DFT+$U$+$V$ calculations are uploaded to a public repository~\cite{DATA}.\\

\section{RESULTS AND DISCUSSIONS}
\subsection{Electronic structures of $ABX_3$-type halide perovskites}
It is common to apply Hubbard corrections to strongly localized $d$ and $f$ electrons~\cite{May20p165117,Gopal15p245202,Yu20p180}, though recent investigations highlighted the importance of applying $U$ corrections to also $s$ and $p$ states~\cite{Tancogne-Dejean20p155117,Lee20p043410, Huang20p165157, Ke21p3387}. In practice, the construction of the Hubbard manifold, namely the orbitals on which the Hubbard Hamiltonian will act, is largely based on empirical rules or intuition. Here, taking CsSnI$_3$ as an example, we compute Hubbard parameters with ACBN0 for 
multiple orbitals at the same time: Sn-5$s$, Sn-5$p$, I-$5s$, I-$5p$, Cs-$6s$, and Cs-$5p$. The self-consistent $U$ values are reported in Table~\ref{UVvales}. The magnitude of $U$ in ACBN0 depends on the degree of localization of the Bloch states because of the introduction of renormalized density matrix, which makes sure that ACBN0 reduces to PBE for delocalized states that are already properly described by PBE~\cite{Agapito15p011006}. This is indeed the case for Cs-$6s$ and Sn-$5p$ states, both have nearly zero $U$ values. In comparison, the onsite $U$ for the Sn-$5s$ states in CsSnI$_3$ acquires a non-negligible magnitude of 2.37~eV, an indication of appreciable electronic correlation effects of lone-pair $5s$ electrons of Sn. 
These results demonstrate that a self-consistent approach that naturally picks up important orbitals to construct the Hubbard manifold is capable of reducing the bias due to an ``ad hoc"  orbital selection. 

\begin{figure}[]
\centering
\includegraphics[width=1\textwidth]{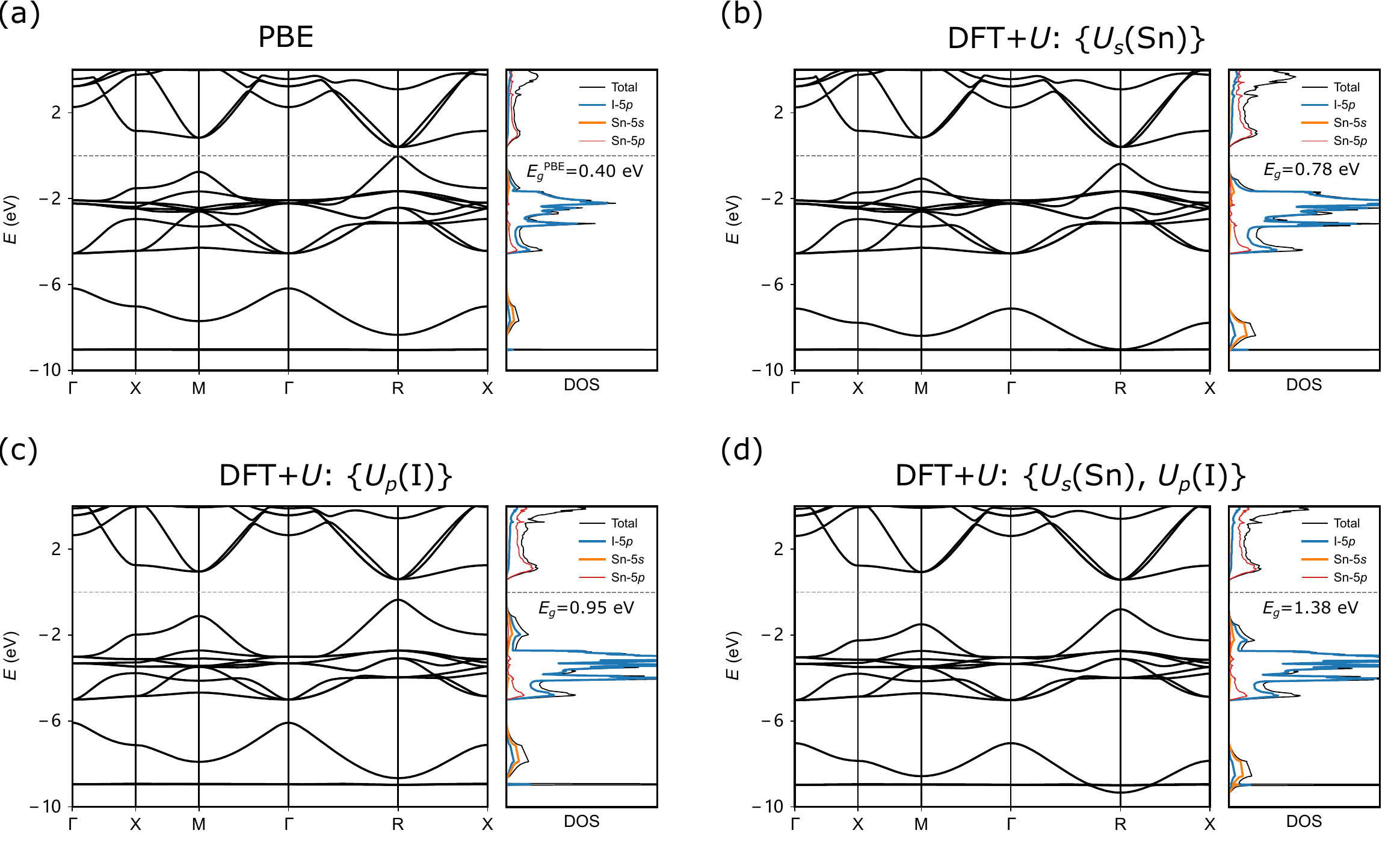}
\caption{Diagnostic analysis of Hubbard $U$ corrections in CsSnI$_3$. Comparison between the band structures and projected density of states (DOS) computed with (a) PBE, (b) ACBN0 with Hubbard $U$ applied to Sn-$5s$ states, (c) ACBN0 with Hubbard $U$ applied to I-$5p$ states, and (d) ACBN0 with $U$ corrections applied to both Sn-$5s$ and I-$5p$ states. $U$ values are reported in Table~\ref{UVvales}. All band structures have the core energies aligned and use the same absolute energy as the Fermi level to show the band shifting resulting from $U$ corrections.}
\label{CsSnI3_U+DOS}
\end{figure}

A detailed analysis is performed to diagnose the effects of Hubbard $U$ on the electronic structure of CsSnI$_3$ by comparing the results obtained with Hubbard manifolds of different constituent local states. The PBE band structure and projected density of states (PDOS) are shown in Fig.~\ref{CsSnI3_U+DOS}a, revealing that the valance-band edge is predominantly of I-$5p$ character with small Sn-$5s$ hybridization while the conduction-band edge takes almost exclusively a Sn-$5p$ character. The isolated single band between $-6$ and $-8$ eV is from Sn-$5s$ and I-$5p$ states. PBE gives a direct band gap of 0.40~eV at $\Gamma$, much lower than the experimental value of 1.3~eV. After applying a Hubbard $U$ correction of 2.37 eV to Sn-$5s$ states (denoted as DFT+$U_s$(Sn)), the band gap increases to 0.78~eV, due to a rigid downshift of the top of the valence manifold (Fig.~\ref{CsSnI3_U+DOS}b). Given the small contribution of Sn-$5s$ states to the valance-band edge, the improved band-gap prediction resulting from $U_s$(Sn) seems puzzling. We find that the use of $U_s$(Sn) downshifts the single band between $-6$ and $-8$ eV such that the $s$-$p$ repulsion between this low-lying band of substantial Sn-$5s$ character and the I-$5p$ bands that dominate the valence-band edge is reduced, eventually leading to the downshift of the valance band maximum (VBM). Applying an on-site Coulomb potential to $5p$ electrons on I sites alone (denoted as DFT+$U_p$(I)) also increases the band gap to 0.95~eV, ascribed to the downshift of I-$5p$ bands (Fig.~\ref{CsSnI3_U+DOS}c). Though DFT+$U_s$(Sn) and DFT+$U_p$(I) give comparable band gap values (0.78 versus 0.95 eV), their PDOS spectra are notably different. Specifically, the PDOS between $-6$ and 0 eV obtained with DFT+$U_p$(I) splits into two regions with a minor peak at $-2$ eV; such splitting is absent in the PDOS spectrum computed with DFT+$U_s$(Sn). The change in PDOS near the Fermi level ($E_F$) will likely impact the electronic  transport properties such as carrier mobility. Finally, the concomitant use of $U_s$(Sn) and $U_p$(I) increases the band gap to 1.38~eV, agreeing well with the experimental value of 1.3~eV. Because bands of Cs-$5p$ and I-$5s$ characters are low-dispersion deep levels, applying $U$ corrections to these semicore states have little impacts on states near $E_F$ and the band gap.

\begin{figure}[]
\centering
\includegraphics[width=0.6\textwidth]{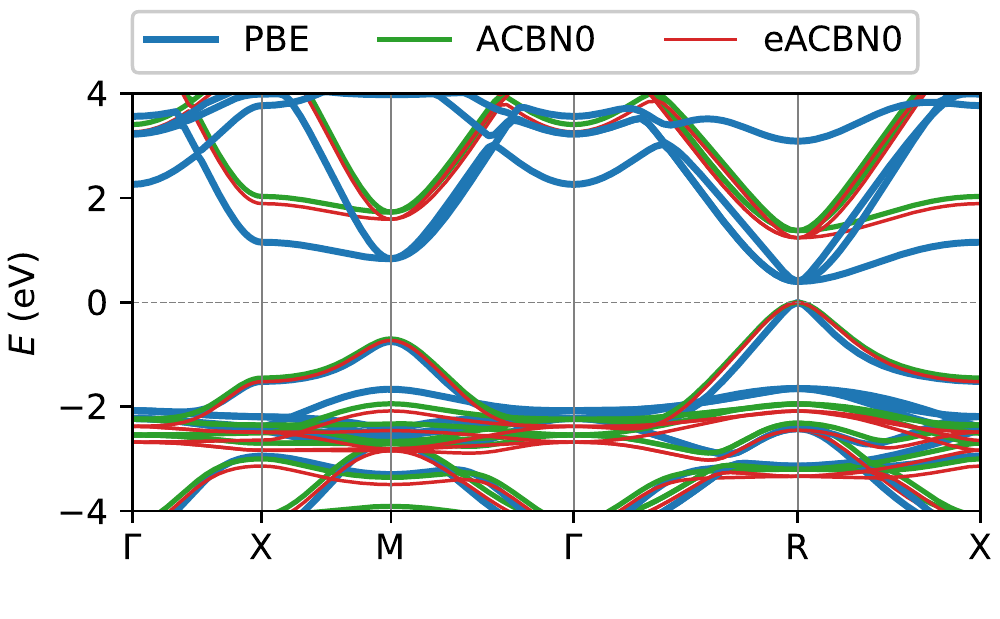}
\caption{Comparison of the band structures of CsSnI$_3$ obtained with PBE, ACBN0, and eACBN0, respectively. The valence band maximum is set as the Fermi level. }
\label{CsSnI3_all}
\end{figure}

We further consider intersite Hubbard interactions between nearest-neighboring Hubbard sites with $V$ parameters computed self-consistently with eACBN0. It is noted that eACBN0 also evaluates all on-site Hubbard parameters on-the-fly, and the $U$ values obtained with ACBN0 and eACBN0 are comparable (Table~\ref{UVvales}). The self-consistent $V$ parameter corresponding to the hybridization between valence $5s$ and $5p$ electrons of nearest-neighboring Sn and I atoms is $\approx$1.6~eV. Interestingly, as illustrated in Fig.~\ref{CsSnI3_all}, the inclusion of this intersite $V$ correction does not alter the band gap nor the band dispersion: ACBN0 and eACBN0 band structures match closely with each other. 
As we will discuss in detail below, the electronic structure of CsSnI$_3$ is insensitive to Hubbard $V$ corrections due to the antibonding nature of VBM. 

\begin{figure}[]
\centering
\includegraphics[width=1\textwidth]{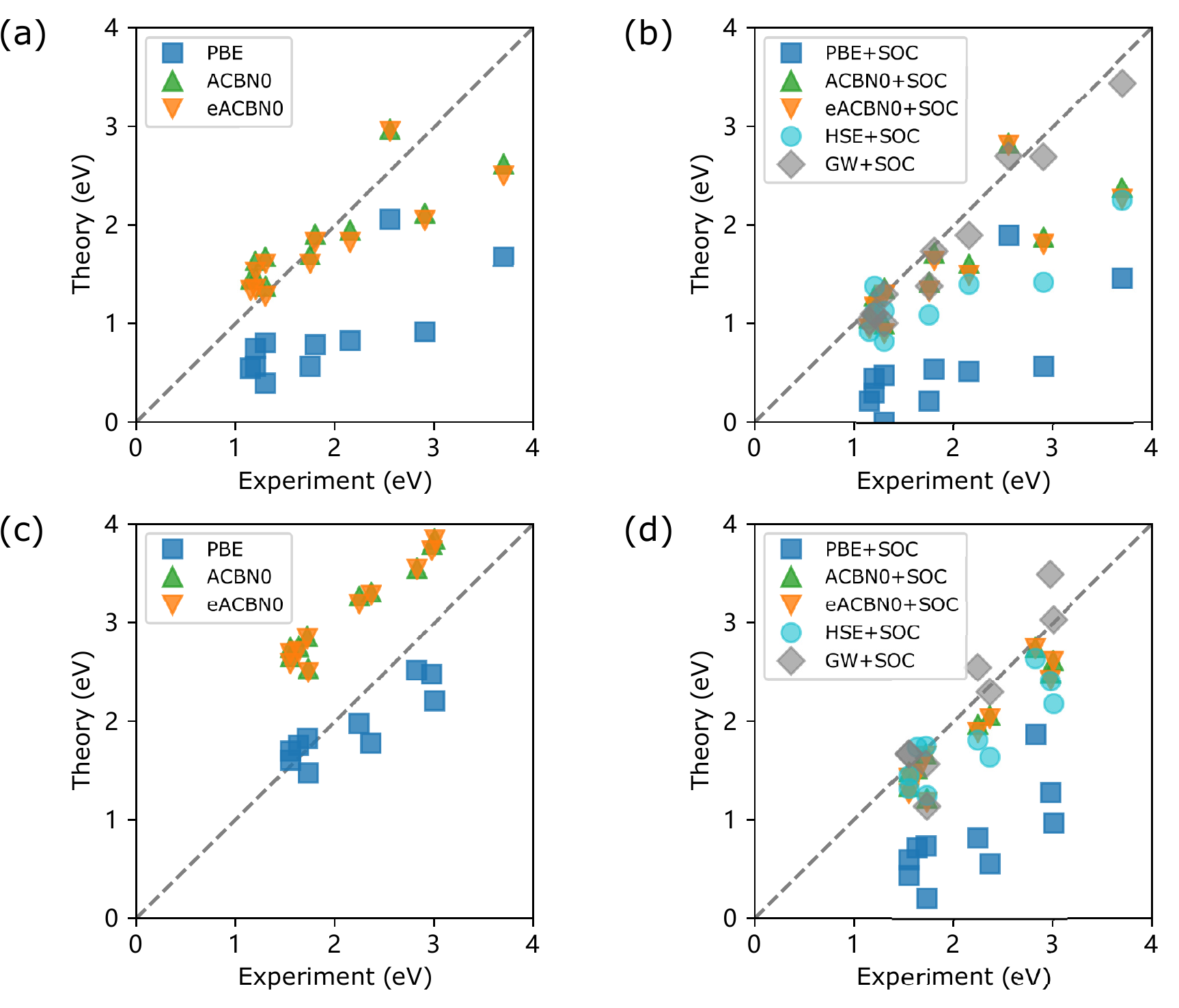}
\caption{Comparison of experimental and theoretical band gaps computed with different methods for Sn-based HPs (a) without and (b) with SOC, and for Pb-based HPs (c) without and (d) with SOC.}
\label{ABX3_all}
\end{figure}

A series of calculations are performed to explore the accuracy of ACBN0 and eACBN0 methods for a list of $ABX_3$ HPs and their polymorphs ($\alpha=$ cubic, $\gamma=$ orthorhombic, and $t=$ tetragonal) with known experimental band gaps. Based on above detailed investigations on CsSnI$_3$, the Hubbard $U$ corrections are applied $B$-site $s$-states and $X$-site $p$-states. The states of $A$-site atoms contribute little to band edges, thus being excluded from the Hubbard manifold. The intersite $V$ between $s$ electrons on $B$ sites and $p$ electrons centered on nearest-neighboring $X$ sites is considered in eACBN0. The band gaps are tabulated in Table~\ref{ABX3} and plotted in Fig.~\ref{ABX3_all}

For Sn-based HPs, PBE, with a mean absolute error (MAE) of 1.01~eV, substantially underestimates the band gap values with reference to experimental results (Fig.~\ref{ABX3_all}a). Both ACBN0 and eABNC0 improve the band gap predictions upon PBE, and their MAEs are 0.37 and 0.36~eV, respectively. The nearly identical MAEs of ACBN0 and eACBN0 indicate the intersite $V$ interactions are not essential in these HPs. We note that this refects a good feature of eACBN0: {\em it reduces to ACBN0 when the electronic structure is already properly described without including intersite $V$.}
Additionally, the inclusion of SOC only reduces the band gap slightly for Sn-based HPs (Fig.~\ref{ABX3_all}b): the average band gap reduction is 0.32~eV for HPs containing I, 0.27~eV for Br, and 0.24~eV for Cl. In general, both ACBN0+SOC (MAE = 0.38 eV) and eACBN0+SOC (MAE = 0.44 eV) are more accurate than HSE+SOC (MAE = 0.61 eV), while $GW$+SOC appears to be the most reliable (MAE = 0.18 eV).

In the case of Pb-based HPs, PBE achieves a good agreement with experimental data due to fortuitous error cancellations in the absence of SOC. As expected, the band gap values obtained with PBE+SOC become much smaller and deviate significantly from experimental values (Fig.~\ref{ABX3_all}d), leading to a large MAE of 1.34~eV. The band gap reduction due to SOC is 1.11~eV. In comparison, ACBN0+SOC and eACBN0+SOC using self-consistent Hubbard parameters all improve upon PBE+SOC and demonstrate satisfying predictive power, a MAE of 0.24~eV for the former and 0.29~eV for the latter. In summary, it is important to include SOC for HPs containing Pb, and both ACBN0+SOC and eACBN0+SOC have consistent accuracy for {\em all} studied $ABX_3$-type HPs but are much more affordable than HSE+SOC and $GW$+SOC.

\subsection{Electronic structure of $A_2BX_6$}
As a typical VODP, Cs$_2$SnI$_6$ has Sn adopting a formal oxidation state of +4, which has been used to explain its enhanced stability relative to CsSnI$_3$~\cite{Lee14p15379}. However, Xiao~\etal\cite{Xiao15p1250} pointed out that the Sn ion in Cs$_2$SnI$_6$ also adopts a +2 oxidation state where [SnI$_6$]$^{2-}$ is better understood as \{Sn$^{2+}$ + [I$_6^{6-}L_2^{2+}$]$^{4-}$\} with $L_2^{2+}$ representing two holes ($L^+$) localized on ligand atoms. Dalpian \etal\cite{Dalpian17p025401} 
used the mechanism of ``self-regulating response" (SRR)~\cite{Raebiger08p763} to explain the relationship between formal oxidation state and the actual charge density residing on the Sn atom: the holes introduced by $B$-site vacancies are absorbed by the whole system through self-consistent metal-ligand rehybridization. Specifically, DFT calculations revealed that the Sn atom in Cs$_2$SnI$_6$ lose 0.6 electrons of 5$s$ character but gain 0.2 $5p$ electrons such that the charge density residing on Sn is almost unchanged compared to that in CsSnI$_3$.
Table~\ref{UVvales} reports the self-consistent Hubbard parameters for Cs$_2$SnI$_6$ computed with ACBN0. The Hubbard $U$  for Sn-$5s$ in Cs$_2$SnI$_6$ is 1.59~eV, slightly smaller than $U_s$(Sn) of 2.37~eV in CsSnI$_3$. This supports the SRR mechanism since the magnitude of self-consistent $U$ in ACBN0 is proportional to the occupancy of the localized orbital; if Sn loses all $5s$ electrons and becomes +4, $U_s$(Sn) would be nearly zero. Therefore, the self-consistent onsite $U$ could serve as a direct measurement of the local charge state. 

\begin{figure}[]
\centering
\includegraphics[width=1\textwidth]{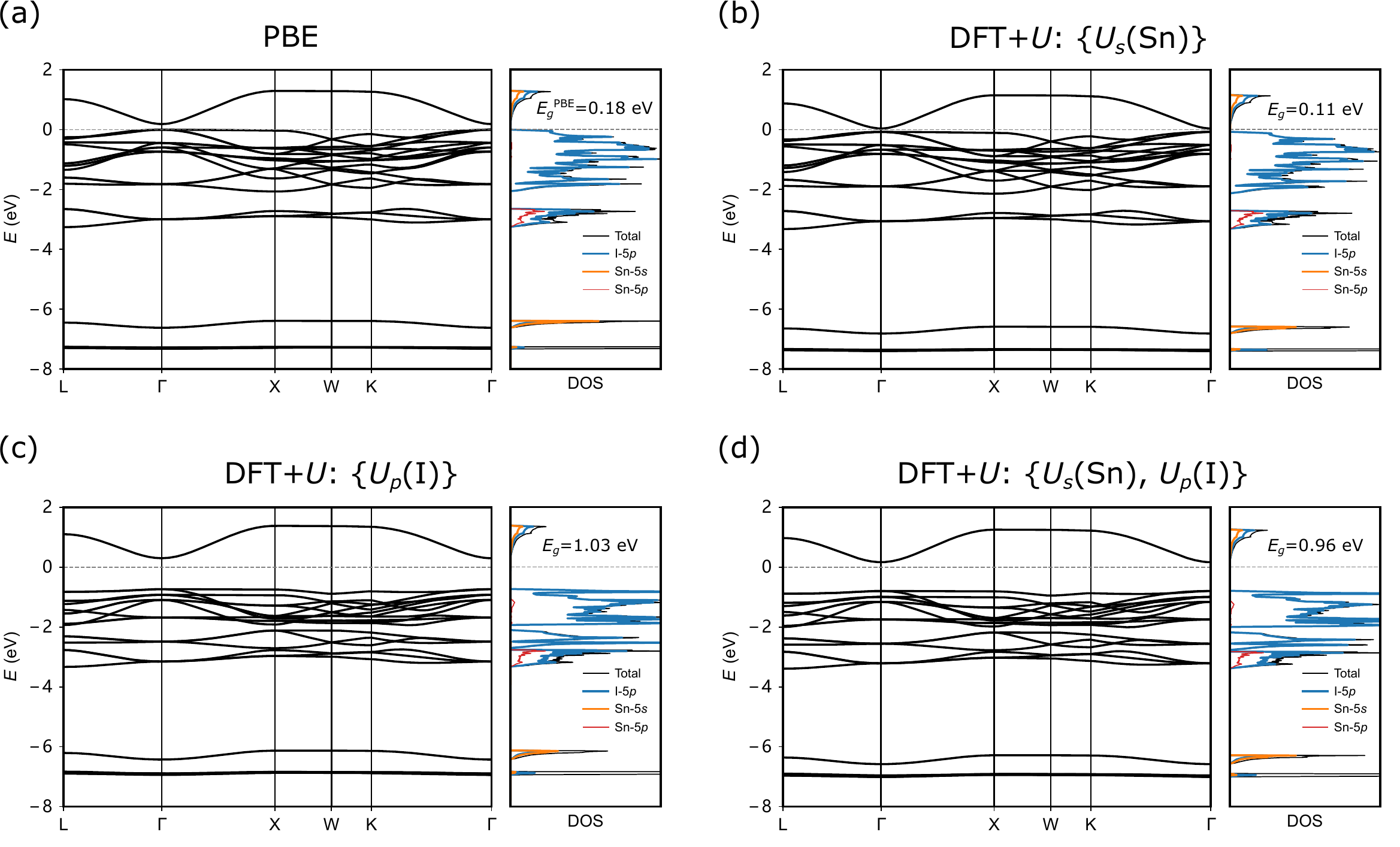}
\caption{Diagnostic analysis of Hubbard $U$ corrections in Cs$_2$SnI$_6$. Comparison between the band structures and projected density of states (DOS) computed with (a) PBE, (b) ACBN0 with Hubbard $U$ applied to Sn-$5s$ states, (c) ACBN0 with Hubbard $U$ applied to I-$5p$ states, and (d) ACBN0 with $U$ corrections applied to both Sn-$5s$ and I-$5p$ states. $U$ values are taken from Table~\ref{UVvales}. All band structures have the core energies aligned and use the same absolute energy as the Fermi level to show the band shifting resulting from $U$ corrections.}
\label{Cs2SnI6_U+DOS}
\end{figure}

An important consequence of metal-ligand rehybridization in Cs$_2$SnI$_6$ is that Sn-$5s$ now has no substantial contribution to the VBM but becomes important at the CBM, as shown in Fig.~\ref{Cs2SnI6_U+DOS}a. 
We perform a diagnostic analysis to comprehend the impacts of onsite $U$ corrections applied to different local orbitals using self-consistent $U$ determined from ACBN0.
In reference to a PBE band gap of 0.18~eV, applying the Hubbard $U$ to Sn-$5s$ states actually pushes down the CBM and decreases the band gap slightly to 0.11 eV (Fig.~\ref{Cs2SnI6_U+DOS}b). Because I-$5p$ states dominate the top of the valence-band manifold, the inclusion of $U_p$(I) downshifts the valance-band edge (Fig.~\ref{Cs2SnI6_U+DOS}c) and gives a band gap of 1.03~eV.
The concomitant use of $U_s$(Sn) and $U_p$(I) yields a band gap of 0.96~eV, agreeing reasonably well with the experimental value of 1.26~eV. 
It is noted that in some previous studies, the percentage ($\alpha$) of Hartree-Fock exact exchange in HSE is adjusted to reproduce the experimental band gap~\cite{Xiao15p1250,Ju18p297}. For Cs$_2$SnI$_6$, HSE with $\alpha=0.25$ predicts a band gap of 0.93~eV, while HSE with $\alpha=0.34$ gives a band gap of 1.26~eV. In contrary, ACBN0 is fully {\em ab initio} without tuning parameters. The band gap of Cs$_2$SnI$_6$ is  found to be insensitive to SOC; for example, PBE and PBE+SOC predicts a band gap of 0.18 and 0.13~eV, respectively. 

\begin{figure}[]
\centering
\includegraphics[width=0.6\textwidth]{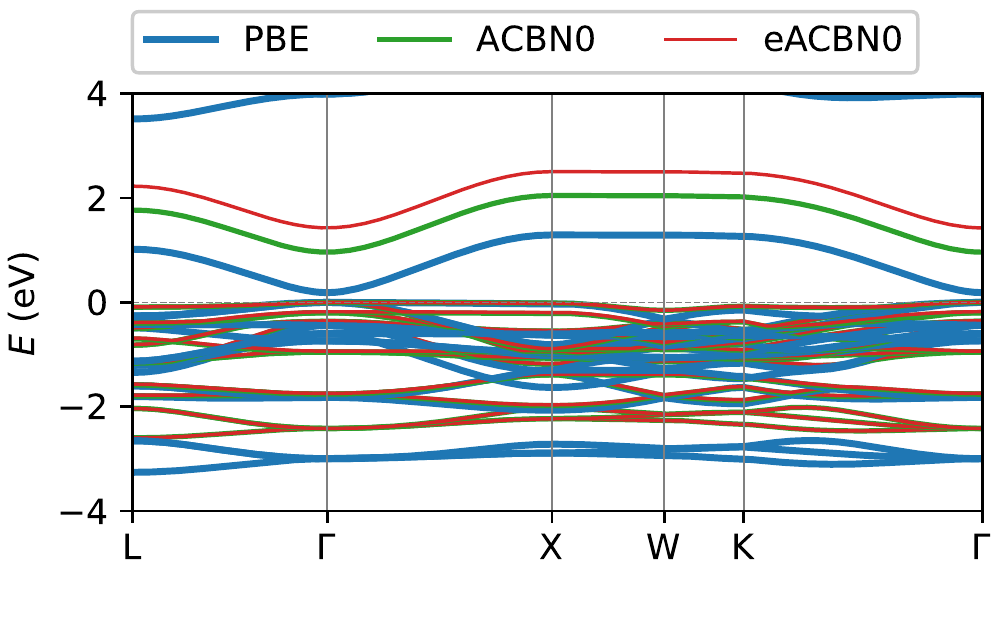}
\caption{Comparison of the band structures of Cs$_2$SnI$_6$ obtained with PBE, ACBN0, and eACBN0, respectively. The valence band maximum is set as the Fermi level.}
\label{Cs2SnI6_all}
\end{figure}

Interestingly, unlike $ABX_3$ HPs that have band gap values insensitive to $V$ corrections, the band gap of Cs$_2$SnI$_6$ increases to 1.42~eV when the intersite Hubbard $V$ interactions between nearest neighboring Sn-$5s$ and I-$5p$ orbitals are included. We find that ACBN0 and eACBN0 band structures have similar band dispersions, and the effect of Hubbard $V$ corrections in eACBN0 is mainly manifested as a more pronounced upshift of the conduction-band manifold (Fig.~\ref{Cs2SnI6_all}). 
 As will be detailed in below, the band gap of Cs$_2$SnI$_6$ opens up monotonically with increasing magnitude of $V$.

\begin{figure}[]
\centering
\includegraphics[width=1\textwidth]{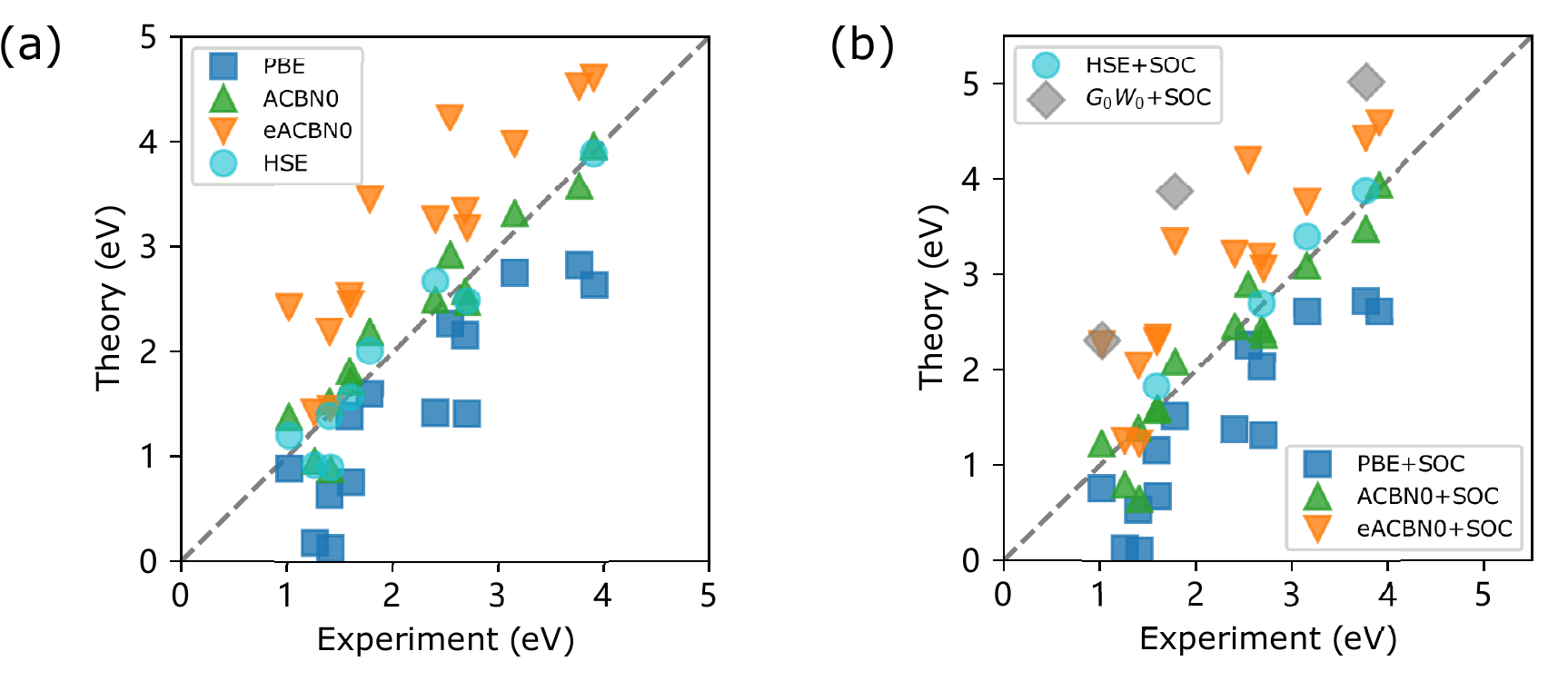}
\caption{Comparison of experimental and theoretical band gaps computed with different methods for $A_2BX_6$ VODPs (a) without and (b) with SOC.}
\label{A2BX6_all}
\end{figure}

Figure~\ref{A2BX6_all} and Table~\ref{A2BX6} compare the PBE, ACBN0, and eACBN0 band gaps with HSE values taken from literature for a few VODPs with known experimental results. The consideration of SOC only slightly reduces the band gap by 0.19~eV for compounds containing I, 0.12~eV for Br, and negligible for Cl.
We find that ACBN0+SOC with a MAE of 0.22~eV is much more accurate than PBE+SOC with a MAE of 0.81~eV, and is comparable with HSE that has a MAE of 0.19~eV. 
Unexpectedly, eACBN0+SOC turns out to be less accurate (MAE = 0.75 eV) and often overestimates the band gap. Despite this seemingly worsened performance, we find that the band gaps predicted by eACBN0 are similar to those obtained with the non-self-consistent many-body $GW$ method, $G_0W_0$, that accounts for dynamical screening. Specifically, $G_0W_0$+SOC predicts the band gaps of Cs$_2$TiI$_6$, Cs$_2$ZrI$_6$, Cs$_2$TiBr$_6$, and Cs$_2$ZrBr$_6$ as 2.31, 3.32, 3.87, and 5.02~eV~\cite{Cucco21p181903}, and eACBN0+SOC gives  2.27, 3.05, 3.35, and 4.43~eV, respectively, all higher than HSE values of 1.20, 2.58, 2.01, and 3.88~eV and available experimental data.
Previous studies showed that the $GW$/$G_0W_0$ approximation yields particularly large errors for materials exhibiting shallow $d$ states, mostly due to the neglect of the attractive
interaction between electrons and holes~\cite{Schilfgaarde06p226402,Shishkin07p246403,Sponza13p235102,Chiodo10p045207}. Given the presence of ligand holes in VODPs, the observed band gap overestimation of eACBN0 is not suprising. 
Therefore, a possible remedy is to include electron-hole interactions in eACBN0 calculations (if possible). 
Nevertheless, we argue that eACBN0 is principally a more reliable method with accuracy on par with $G_0W_0$, but in the case of VODPs, the ``trick" of error cancellation in ACBN0 (and HSE) somehow works out better than eACBN0 and $G_0W_0$. 
It was also suggested that because VODPs are not defect tolerant such that the presence of in-gap defect states could strongly affect the measurement that gives a experimental band gap lower than the intrinsic forbidden gap. This could also explain the ``overestimation" of eACBN0 for VODPs.

\subsection{Effects of Hubbard $V$ interactions in $ABX_3$ and $A_2BX_6$ HPs}
\begin{figure}[]
\centering
\includegraphics[width=0.4\textwidth]{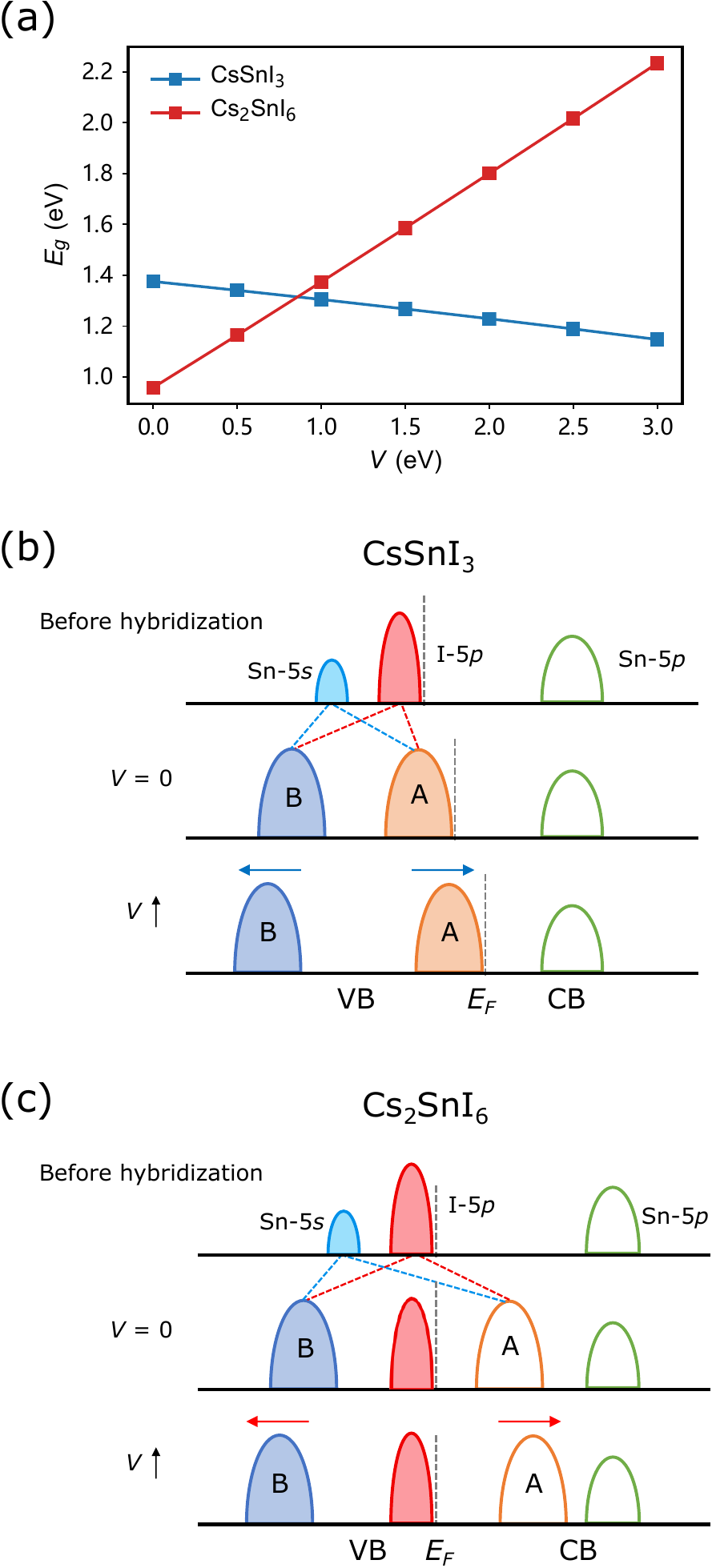}
\caption{Effects of Hubbard $V$ interactions on the band gaps of CsSnI$_3$ and Cs$_2$SnI$_6$. (a) Evolution of the band gaps as a function of Hubbard $V$. Schematic diagram of the Sn-$5s$ and I-$5p$ energy levels in (b) CsSnI$_3$ and (c) Cs$_2$SnI$_6$ before (top panel) and after (middle and bottom panels) hybridization. The antibonding and bonding bands are labeled as A and B, respectively. In CsSnI$_3$, the band gap is mostly determined by the energy difference between the occupied antibonding states and the empty Sn-$5p$ states. A larger $V$ that enhances the $s$-$p$ hybridization will effectively push up the energy of antibonding states thus reducing the band gap. In Cs$_2$SnI$_6$, the VBM is dominated by states of I-$5p$ characters that is insensitive to $V$. A larger $V$ will push up the antibonding states and increase the band gap. 
}
\label{VinHPs}
\end{figure}

We perform a set of model calculations to understand the effects of Hubbard $V$ interactions on the band gaps of CsSnI$_3$ and Cs$_2$SnI$_6$ by gradually increasing the magnitude of $V$ parameters between nearest-neighboring Sn-$5s$ and I-$5p$ orbitals. As shown in Fig.~\ref{VinHPs}a, the band gap of CsSnI$_3$ decreases linearly with increasing $V$ while the band gap of Cs$_2$SnI$_6$ increases monotonically. Such opposite $V$-dependence of the band gap in these two different types of HPs can be understood using the energy level diagrams (Fig.~\ref{VinHPs}b-c). Before the hybridization, I-$5p$ states have higher energy than Sn-$5s$ states. The hybridization leads to the splitting between the bonding and antibonding bands. This is the case for both  CsSnI$_3$ and Cs$_2$SnI$_6$. In CsSnI$_3$, both bonding and antibonding states are occupied and the VBM is composed of antibonding states; the band gap is determined by the energy difference between the occupied antibonding states and the empty Sn-$5p$ states. The intersite $V$ parameter measures the strength of hybridization between neighboring Hubbard sites. Therefore, a stronger hybridization between Sn-$5s$ and I-$5p$ will increase the bonding-antibonding energy splitting that effectively pushes up the VBM and leads to a smaller band gap (Fig.~\ref{VinHPs}b). In comparison, previous studies~\cite{Xiao15p1250,Dalpian17p025401} have confirmed that the VBM of Cs$_2$SnI$_6$ is formed by the antibonding states between I-$5p$ orbtials that are nonbonding to Sn, whereas the CBM is composed of the antibonding states arising from the hybridization between Sn-$5s$ orbitals and [I$_6$] $a_{1g}$ molecular orbitals. It is then not surprising that the VBM is not sensitive to $V$ while a larger $V$ will push the CBM up and increase the band gap (Fig.~\ref{VinHPs}c). The enhanced stability of Cs$_2$SnI$_6$ relative to CsSnI$_3$ can thus be understood as the depletion of antibonding states rather than Sn ions acquiring a higher oxidation state of +4. 
Different from CsSnI$_3$ that has a small (negative) slope of the $E_g$-$V$ line, the band gap of Cs$_2$SnI$_6$ is more sensitive to Hubbard $V$, indicating that the Sn-I bonds in this VOHP are much more covalent. 

\subsection{Electronic structures of BDHPs}
\begin{figure}[]
\centering
\includegraphics[width=1\textwidth]{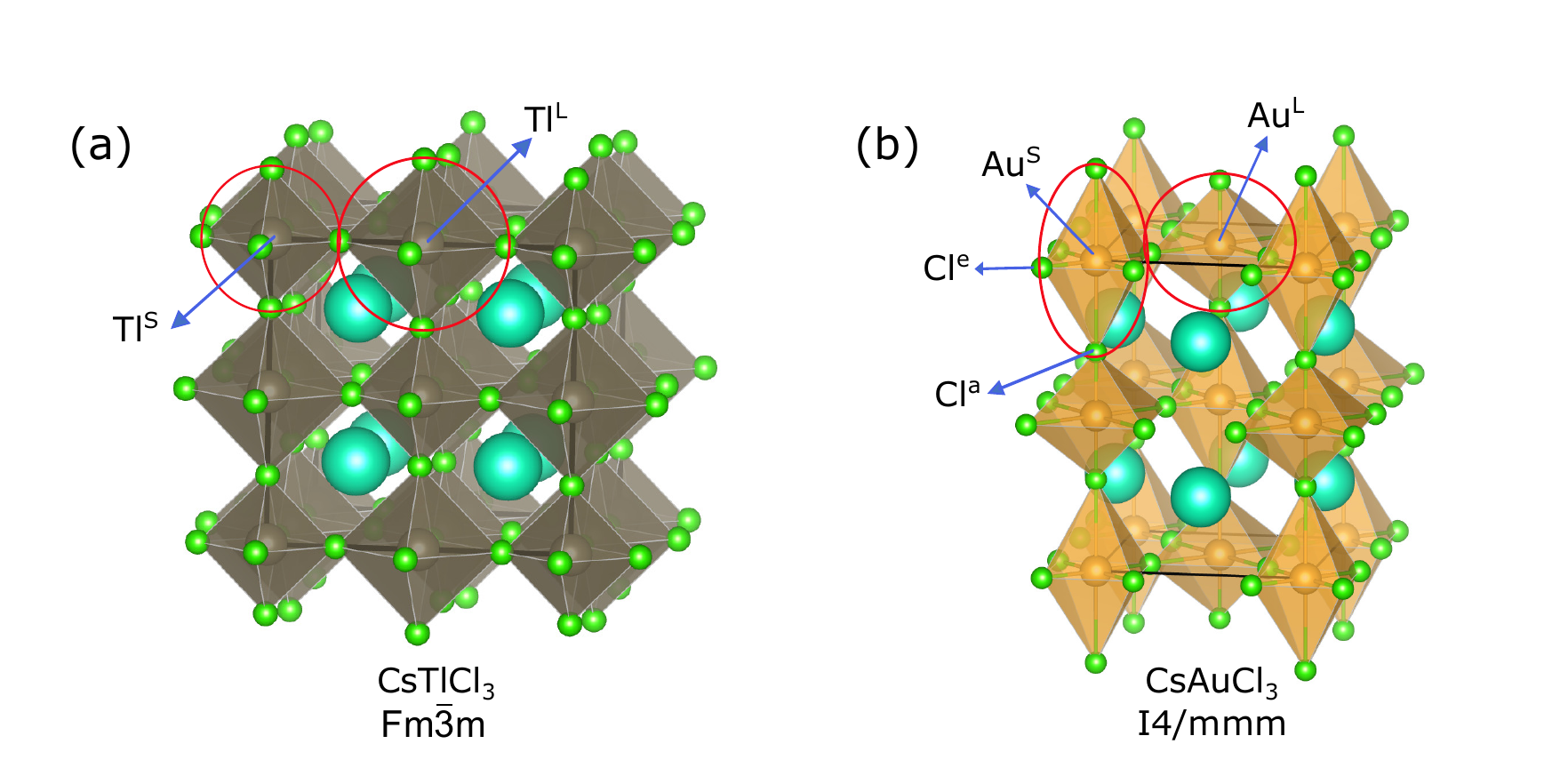}
\caption{Crystal structure of (a) cubic CsTlCl$_3$ in the space group of $Fm\bar{3}m$ and (b) tetragonal CsAuCl$_3$ in the space group of $I4/mmm$. Because of the breathing-mode 
distortions of halide octahedra, there are two different local environments (DLEs) associated with the same $B$ element. The Tl atoms in large and small halide cages are labled as Tl$^{\rm L}$ and  Tl$^{\rm S}$, respectively. Same notations are used for Au$^{\rm L}$ and Au$^{\rm S}$. The smaller [Cl$_6$] cage in CsAuCl$_3$ is also strongly elongated along the $c$ axis, and the two symmetry-inequivalent Cl atoms are labeled as Cl$^{\rm a}$ for the the axial site and Cl$^{\rm e}$ for the equatorial site. 
}
\label{BDHPstructure}
\end{figure}

The structures of two typical BDHPs, CsTlCl$_3$ and  CsAuCl$_3$, are shown in Fig.~\ref{BDHPstructure}. Both compounds have breathing-mode 
distortions of the halide octahedra that leads to two different local environments (DLEs) associated with the same $B$ element. In the unit cell of CsTlCl$_3$ (space group $Fm\bar{3}m$), the TlF$_6$ cages are alternately dilated and contracted isotropically, with Tl-Cl bonds in each individual cage being of the same length.  The crystal structure of CsAuCl$_3$ has alternately compressed and elongated [AlCl$_6$] cages along the $c$ axis. Following the suggestions in ref.~\cite{Benam21p195141,Dalpian18p075135}, we label the $B$-site cations based on the bond geometry instead of formal oxidation state to avoid the misinterpretation of charge ordering/disproportion, that is, $B^{\rm L}$ and $B^{\rm S}$ for a $B$ atom in a large and small octahedron, respectively.  The smaller [Cl$_6$] cage in CsAuCl$_3$ is also strongly elongated along the $c$ axis, and the two symmetry-inequivalent Cl atoms are labeled as Cl$^{\rm a}$ for the the axial site and Cl$^{\rm e}$ for the equatorial site (Fig.~\ref{BDHPstructure}b). 

We compute the band gaps of CsTlCl$_3$ and CsAuCl$_3$ using PBE, ACBN0, and eACBN0, and compare them with HSE and/or experimental values if available (Table~\ref{BDHPgaps}). 
For CsTlCl$_3$, ACBN0 improves upon PBE, yielding a band gap of 1.50 eV, and eACBN0 further increases the band gap to 1.89~eV that is higher than HSE value of 1.3~eV but compares favorably with the experimental value of 2.5~eV. With regard to CsTlF$_3$, we observe a similar trend: eACBN0 band gap of 3.10~eV is larger than ACBN0 band gap of 2.64~eV but smaller than HSE value of 3.9~eV. Interestingly for CsAuCl$_3$, PBE and ACBN0 predict similar band gaps (0.95 versus 0.90~eV), while eACBN0 gives a band gap of 1.37~eV that is lower than HSE value of 1.51~eV and experimental value of 2.04~eV. Overall, it is critical to take into account the hybridizations between $B$-site cations and halides to achieve better descriptions of the electronic structures of BDHPs. 

\begin{figure}[]
\centering
\includegraphics[width=1\textwidth]{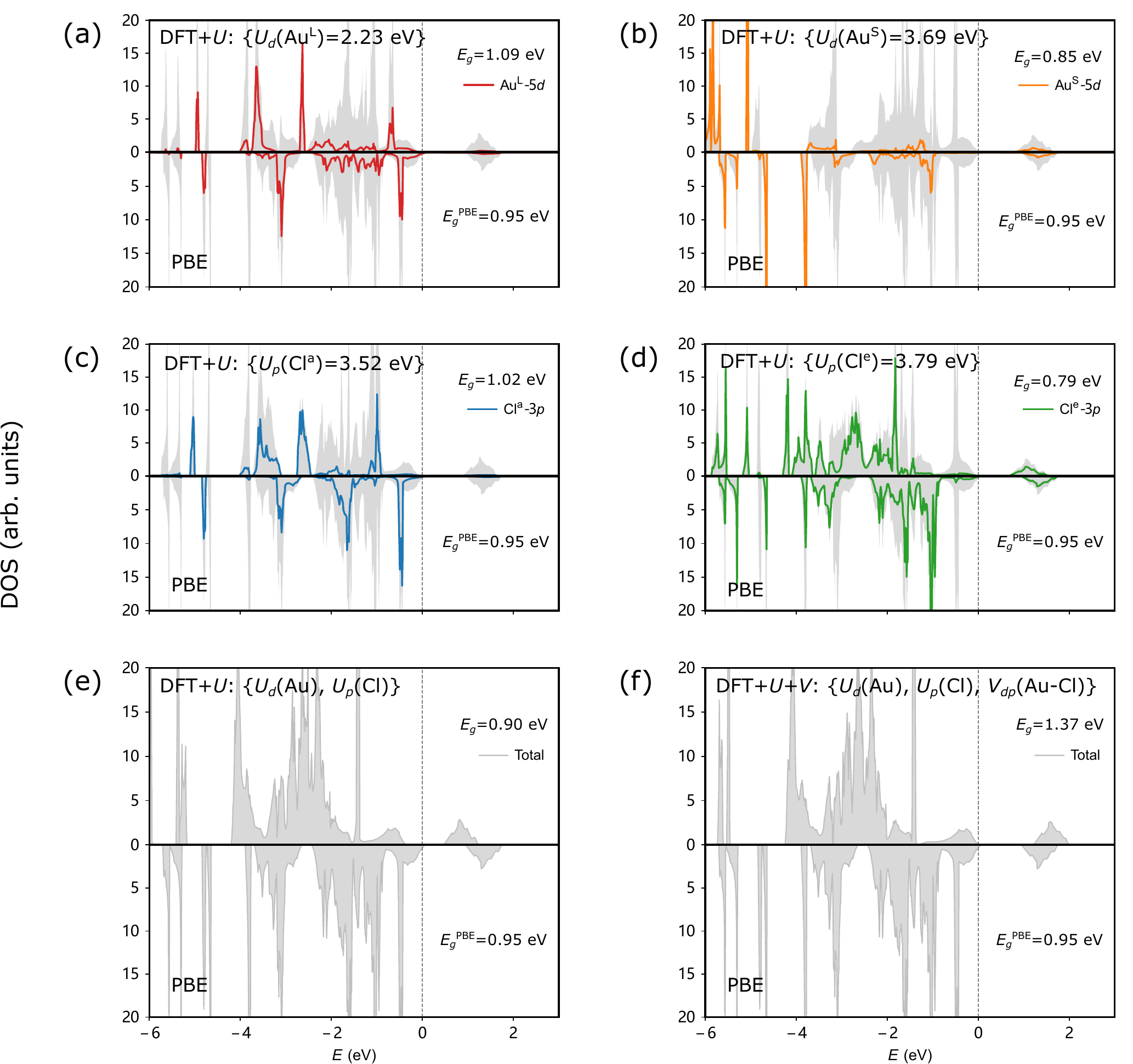}
\caption{
Diagnostic analysis of Hubbard $U$ and $V$ corrections in CsAuI$_3$. Comparison of density of states computed with PBE and those computed with self-consistent Hubbard $U$ corrections applied to (a) Au$^{\rm L}$-$5d$ states, (b) Au$^{\rm S}$-$5d$ states, (c) Cl$^{\rm a}$-$3p$ states, (d) Cl$^{\rm e}$-$3p$ states, and (e) both Au-$5d$ and Cl-$3p$ states. Both $U$ and $V$ correlations are included in (f) that clearly shows an upshift of the CBM in eACBN0.
}
\label{CsAuCl3_pDOS}
\end{figure}

The observation that ACBN0 and PBE predict nearly the same band gap values for CsAuCl$_3$ is puzzling. We perform a detailed analysis by gradually introducing self-consistent $U$ corrections to different orbitals. It is found that applying $U_{d}$ of 2.23~eV to Au$^{\rm L}$-5$d$ states slightly increases the band gap to 1.09 eV (Fig.~\ref{CsAuCl3_pDOS}a), mainly due to the downshift of the VBM that has Au$^{\rm L}$-5$d$ characters. In comparison, applying Hubbard $U$ to the $5d$ states of Au$^{\rm S}$ atoms reduces the band gap to 0.85 eV caused by the downshift of the CBM that comprises of Au$^{\rm S}$-$5d$ states (Fig.~\ref{CsAuCl3_pDOS}b). When the Hubbard $U$ corrections are applied to both Au$^{\rm L}$-$5d$ and Au$^{\rm S}$-$5d$ states, their effects on the band gap cancel out each other. Similar canceling effect is also found for Cl$^{\rm a}$-$3p$ and Cl$^{\rm e}$-$3p$ (see Fig.~\ref{CsAuCl3_pDOS}c-d). Consequently, the collective Hubbard $U$ corrections to $5d$-states of Au and $3p$-states of Cl, though strongly affect the DOS spectrum, does not change the band gap appreciably (Fig.~\ref{CsAuCl3_pDOS}e). It is only after the application of intersite $V$ corrections that the band gap increases due to the upshift of the CBM (Fig.~\ref{CsAuCl3_pDOS}f).

\subsection{Effects of Hubbard $V$ interactions in BDHPs}
\begin{figure}[]
\centering
\includegraphics[width=0.4\textwidth]{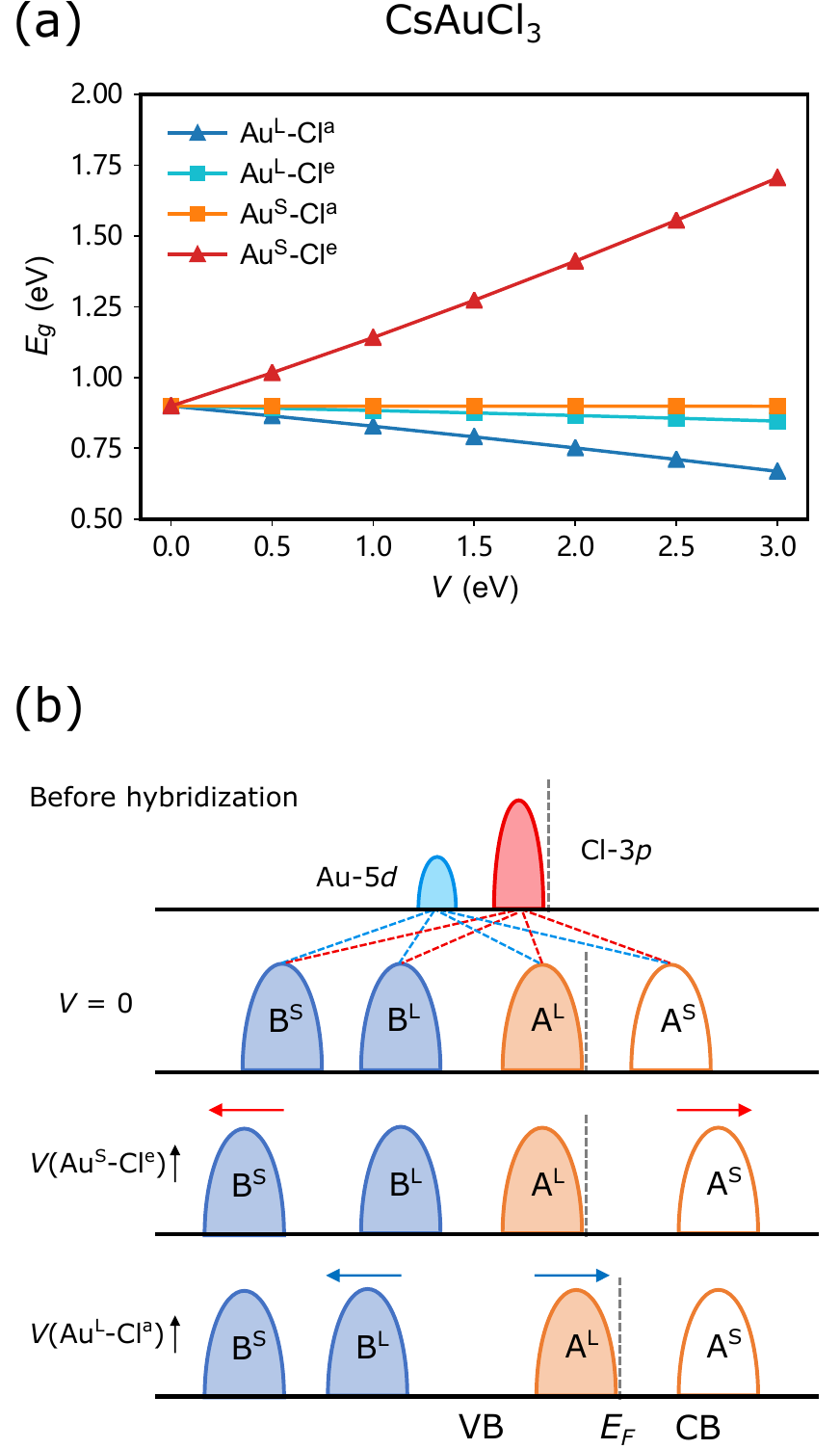}
\caption{Effects of Hubbard $V$ interactions on the band gaps of CsAuCl$_3$. (a) Evolution of the band gaps as a function of Hubbard $V$. There are four nonequivalent Au-Cl bonds in CsAuCl$_3$, corresponding to four $V$ parameters. Au$^{\rm S}$-Cl$^{\rm e}$ and Au$^{\rm L}$-Cl$^{\rm a}$ are shortest two bonds. (b) Schematic diagram of the Au-$5d$ and Cl-$3p$ energy levels before (top panel) and after hybridization. The hybridization between Au-$5d$ and Cl-$3p$ orbitals leads to the splitting between the bonding and antibonding states. The antibonding bands resulted from $d$-$p$ hybridization of the smaller [AuCl$_6$] cage, denoted as A$^{\rm S}$, are higher in energy and are unoccupied, while the antibonding bands labeled as A$^{\rm L}$ contribute to the VBM. A larger $V$(Au$^{\rm S}$-Cl$^{\rm e}$) will push up A$^{\rm S}$ and increase the band gap, while an increase in $V$(Au$^{\rm L}$-Cl$^{\rm a}$) will upshift A$^{\rm L}$ and cause a band gap reduction.
}
\label{VinBDHP}
\end{figure}

The inclusion of the intersite Hubbard V is crucial to properly capture the hybridization
between valence electrons on neighboring atoms in BDHPs. A detailed understanding of the effects of $V$ is therefore worthy of investigations.  
There are four nonequivalent Au-Cl bonds in CsAuCl$_3$ with $r$(Au$^{\rm L}$-Cl$^{\rm a}$) $<$ $r$(Au$^{\rm S}$-Cl$^{\rm e}$) $<$ $r$(Au$^{\rm L}$-Cl$^{\rm e}$) $<$ $r$(Au$^{\rm S}$-Cl$^{\rm a}$), corresponding to  different degrees of hybridizations and four $V$ parameters (Table~\ref{CsAuCl3_bonds}). 
A set of model calculations reveal that the band gap of CsAuCl$_3$ opens up with increasing $V$(Au$^{\rm S}$-Cl$^{\rm e}$), but decreases with increasing $V$(Au$^{\rm L}$-Cl$^{\rm a}$), while being insensitive to both $V$(Au$^{\rm L}$-Cl$^{\rm e}$) and $V$(Au$^{\rm S}$-Cl$^{\rm a}$). That is, the electronic structure is more sensitive to the intersite $V$ paramters of two shorter bonds. The trend can be understood with the diagram illustrated in Fig.~\ref{VinBDHP}. The hybridization between Au-$5d$ and Cl-$3p$ orbitals leads to the splitting between the bonding and antibonding states. Because of the presence of two DLEs and the associated differing strengthes in hybridization, the antibonding states (denoted as A$^{\rm S}$) resulting from Au-$5d$ and Cl-$3p$ orbitals of the smaller [AuCl$_6$] cage  are higher in energy and are unoccupied; the antibonding states (denoted as A$^{\rm L}$) resulting from Au-Cl hybridizations in the larger [AuCl$_6$] cage  are occupied and contribute to the VBM. As a result, a large $V$(Au$^{\rm S}$-Cl$^{\rm e}$) will push up the CBM and increase the band gap, while an increase in $V$(Au$^{\rm L}$-Cl$^{\rm a}$) will upshift the VBM and cause a gap reduction. A similar trend is also found in CsTlF$_3$ and CsTlCl$_3$ as well. For example, CsTlF$_3$ has the band gap increases with increasing $V$(Tl$^{\rm S}$-F).

\begin{figure}[]
\centering
\includegraphics[width=0.6\textwidth]{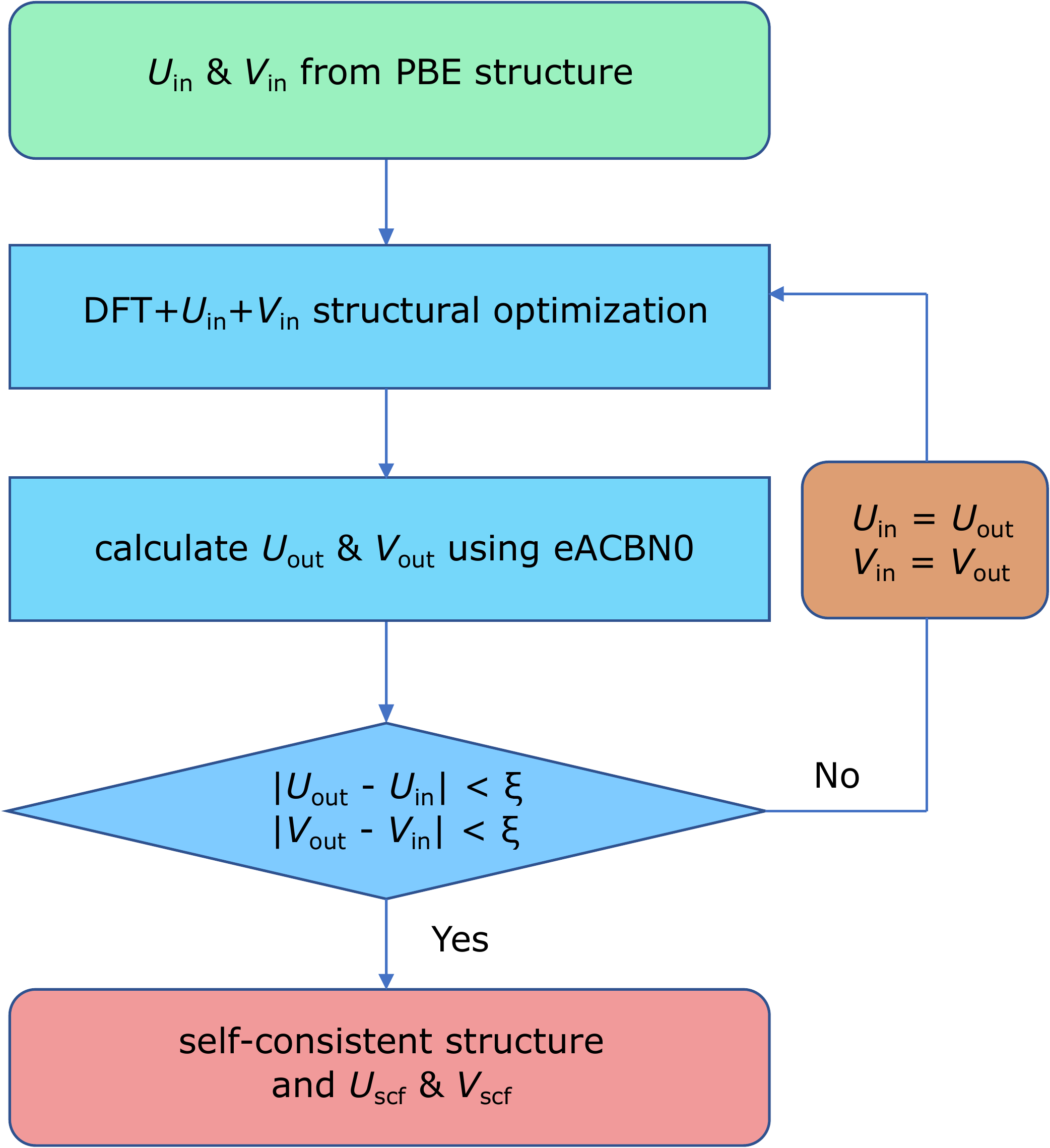}
\caption{Flow chart of sc-eACBN0 that converges both geometry and Hubbard parameters.}
\label{flow chart}
\end{figure}

It is evident that the band gaps of BDHPs depend on the relative hybridization strength between two DLEs. A recent work~\cite{Jang22arxiv} shows that the self-consistent Hubbard $V$ plays a decisive role in describing the coupled charge and lattice degrees of freedom in charge ordered systems such as Ba$_{1-x}$K$_xA$O$_3$ ($A$ = Bi and Sb)~\cite{Jang22arxiv}.
So far, we obtain self-consistent $U$ and $V$ parameters for a given structure optimized with PBE. This then raises a few important questions: is it possible to converge both Hubbard parameters as well as crystal geometry at the same time? How large an impact will this be on the electronic structure? 
To address these questions, we follow a protocol developed in ref.~\cite{Timrov21p045141} that drives the system to the ground state while fully accounting for the changes in the Hubbard parameters. The major difference between our protocol (termed as sc-eACBN0) in Fig.~\ref{flow chart} and that in ref.~\cite{Timrov21p045141} is we use eACBN0 instead of density functional perturbation theory to compute Hubbard parameters self-consistently~\cite{Timrov18p085127}. The band gaps and converged Hubbard parameters are reported in Table~\ref{BDHPgaps} and Table~\ref{UVvales_BDHP}. Interestingly, the band gap prediction of sc-eACBN0 improves greatly over eACBN0. For example, sc-eACBN0 gives a band gap of 2.67~eV for CsTlCl$_3$ and 1.73~eV for CsAuCl$_3$, both in reasonable agreement with experimental values of 2.5 and 2.04~eV, respectively. Structurally, the equilibrium lattice constants and Au-Cl bond lengths obtained with sc-eACBN0 are $\approx$1.7\% larger than the corresponding PBE values (Table~\ref{CsAuCl3_bonds}). Such ``structural dilation" has been previously observed for LiMnPO$_4$ and MnPO$_4$ that DFT+$U$+$V$ predicts larger lattice constants than standard DFT~\cite{Timrov21p045141}.  Overall, we believe sc-eACBN0 is a cost-effective and accurate approach to predict the electronic structures of BDHPs. 

\section{CONCLUSION}
The vast diversity of chemical space enabled by HPs opens up exciting opportunities to obtain novel materials with highly tunable electronic properties for a broad range of applications. 
It is desirable to have a fully {\em ab initio} method to accurately predict the electronic structures of HPs with minimum human bias and affordable computational expense. 
In this work, we investigate three different types of HPs that have drastically different bonding characters using DFT+$U$ and DFT+$U$+$V$ with onsite $U$ and intersite $V$ Hubbard parameters computed self-consistently without ad hoc assumption.  We demonstrate that ACBN0 and eACBN0, a DFT+$U$ method and its extended version with $V$, have improved band gap prediction accuracy over PBE and on par with HSE/GW. Specifically, the finding that the $5s$-electrons of Sn in CsSnI$_3$ and Cs$_2$SnI$_6$ acquire similar on-site Hubbard $U$ values support the mechanism of charge self-regulating response: Sn ions in these two compounds actually have comparable charge states. This highlights the self-consistent $U$ that depends sensitively on  the local atomic environment can serve as a useful indicator of the true charge state.   
Moreover, the simultaneous convergence of both Hubbard parameters and crystal geometry enables a band gap prediction accuracy superior to HSE for bond disproportionated halide perovskites that have complex distortions of the halide octahedra. 
Our work paves the way for future studies of complex compounds beyond HPs containing localized electrons  that often possess intricate competitions between onsite localization and intersite hybridization.
\\

\begin{acknowledgments}
J.Y., T.Z., S.L. acknowledge the supports from Westlake Education Foundation. The computational resource is provided by Westlake HPC Center.
\end{acknowledgments}
{\bf{Competing Interests}} The authors declare no competing financial or non-financial interests.

{\bf{Data Availability}} The data that support the findings of this study are included in this article and are available from the corresponding author upon reasonable request.


\bibliography{main.bbl}

\clearpage
\newpage
\begin{table}[]
\centering
\caption{Self-consistent $U$ and $V$ values (in eV) for different orbitals in CsSnI$_3$ and Cs$_2$SnI$_6$ comptuted with ACBN0 and eACBN0 using the ground-state structures optimized with PBE. $V_{sp}$ is the intersite Hubbard term between nearest-neighboring Sn-$5s$ and I-$5p$; $V_{pp}$ is for intersite interactions between Sn-$5p$ and I-$5p$. Underlined Hubbard parameters are most important for band gap predictions.}
\label{UVvales}
\begin{tabular}{ccccc}
\hline
\hline
\multirow{2}{*}{} & \multicolumn{2}{c}{CsSnI$_3$} & 
\multicolumn{2}{c}{Cs$_2$SnI$_6$}  \\
\cline{2-3}\cline{4-5}
& ACBN0      &eACBN0     & ACBN0     & eACBN0  \\
\hline
$U$(Cs-6$s$)       & 0.01    & 0.01     & 0.10   & 0.10         \\
$U$(Cs-5$p$)       & 5.80    & 5.67     & 6.45   & 6.42        \\
\underline{$U$(Sn-5$s$)}      & 2.37    & 2.26     & 1.59   & 1.05     \\
$U$(Sn-5$p$)       & 0.10    & 0.14     & 0.13   & 0.15         \\
$U$(I-5$s$)        & 9.75    & 9.81     & 9.34   & 9.26      \\
\underline{$U$(I-5$p$)}       & 4.08    & 4.07     & 4.46   & 4.40      \\
\underline{$V_{sp}$}          & -       & 1.60     & -      & 1.18     \\
$V_{pp}$           & -       & 1.57     & -      & 1.67       \\
\hline
\hline
\end{tabular}
\end{table}

\clearpage
\newpage
\begin{table}[]
\centering
\caption{Comparison of the band gaps (in eV) as computed with different methods and as measured in experiments for prototypical $ABX_3$ HPs. The mean absolute error (MAE) of each method is computed for Sn-based HPs and Pb-based HPs separately; MAE$_t$ reflects the accuracy for all $ABX_3$ HPs.}
\label{ABX3}
\begin{tabular}{c|ccc|ccc|ccc}
\hline
\hline
\multirow{3}{*}{$ABX_3$} & \multicolumn{6}{c|}{This Work} & 
\multicolumn{3}{c}{Reference}  \\
\cline{2-7}\cline{8-10}
 &\multicolumn{3}{c|}{without SOC} &  \multicolumn{3}{c|}{with SOC} & \multirow{2}{*}{HSE+SOC} & \multirow{2}{*}{GW+SOC} & \multirow{2}{*}{Expt} \\
 \cline{2-7}
 & PBE      & ACBN0    &eACBN0 &  PBE      & ACBN0    &eACBN0  \\
\hline
$\gamma$-CsSnI$_3$   & 0.81  & 1.68 & 1.61 &0.48  & 1.36  &1.29  &1.13~\cite{Das22p2184}  & 1.3~\cite{Huang13p165203} & 1.3~\cite{Chen12p345}   \\
$Y$-CsSnI3           & 2.06  &2.97 &2.95 &1.90  & 2.83  &2.81  & -      & 2.7~\cite{Huang13p165203}  & 2.55~\cite{Chen12p345}   \\
$\beta$-CsSnBr$_3$   & 0.79  &1.91 &1.83 &0.54  & 1.72  &1.64  & -      & 1.740~\cite{Huang13p165203} & 1.8~\cite{Clark1981p133}   \\
$\alpha$-CsSnI$_3$   & 0.40  &1.38 &1.24 &0.005 & 1.00  &0.86  &0.82~\cite{Das22p2184}  & 1.008~\cite{Huang13p165203} & 1.3~\cite{Dalpian17p025401}  \\
$\alpha$-CsSnBr$_3$  & 0.57  &1.70 &1.61 &0.22  & 1.42  &1.33  &1.09~\cite{Das22p2184}  & 1.382~\cite{Huang13p165203} & 1.75~\cite{Peedikakkandy16p19857} \\
$\alpha$-CsSnCl$_3$  & 0.92  &2.12 &2.05 &0.57  & 1.88  &1.81  &1.42~\cite{Das22p2184}  & 2.693~\cite{Huang13p165203} & 2.9~\cite{Voloshinovskii1994p226}  \\
t-MASnI$_3$   & 0.57   &1.47 &1.35 & 0.30   & 1.15   & 1.03   & 1.11~\cite{Das22p2184}   & 1.10~\cite{Umari14p1} & 1.2~\cite{Stoumpos13p9019}  \\
$\gamma$-MASnI$_3$   &0.75   &1.64 &1.53 &0.45   &1.28   &1.17  &1.38~\cite{Das22p2184}  & -     &1.20~\cite{Parrott16p1321}    \\
$\alpha$-MASnI$_3$   &0.55   &1.45 &1.34 &0.22   &1.06   &0.95  &0.92~\cite{Das22p2184}  &1.03~\cite{Bokdam16p1}  &1.15~\cite{Dang16p3447}  \\
$\alpha$-MASnBr$_3$  &0.83   &1.95 &1.83 &0.52   &1.61   &1.49  &1.4~\cite{Das22p2184}  &1.90(GW)~\cite{Chiarella08p045129} &2.15~\cite{Chiarella08p045129} \\
$\alpha$-MASnCl$_3$  &1.68   &2.62 &2.50 &1.46   &2.38   &2.26  &2.25~\cite{Das22p2184} &3.44(GW)~\cite{Chiarella08p045129} &3.69~\cite{Chiarella08p045129} \\
\bf {MAE} &1.01  &0.37 &0.36 &1.30  &0.38  &0.44  &0.61  & 0.18 \\
\hline
$\gamma$-CsPbI$_3$   & 1.83  &2.86 &2.84 &0.74  & 1.67  &1.65  & 1.75~\cite{Das22p2184} & 1.57~\cite{Sutton18p1787}  & 1.72~\cite{Eperon14p982} \\
$\delta$-CsPbI$_3$   & 2.52  &3.55 &3.54 &1.87  & 2.75  &2.74  & 2.64~\cite{Brgoch14p27721} & -         & 2.82~\cite{Wang19p14501} \\
$\alpha$-CsPbI$_3$   & 1.48  &2.54 &2.50 &0.21  & 1.22  &1.18  & 1.25~\cite{Das22p2184} & 1.14~\cite{Sutton18p1787}  & 1.73~\cite{Eperon14p982} \\
$\alpha$-CsPbBr$_3$  & 1.78  &3.31 &3.28 &0.56  & 2.06  &2.03  & 1.64~\cite{Das22p2184} & 2.30~\cite{Lang15p075102}  & 2.36~\cite{Gesi1975p463} \\
$\alpha$-CsPbCl$_3$  & 2.21  &3.85 &3.84 &0.97  & 2.62  &2.61  & 2.18~\cite{Das22p2184} & 3.03~\cite{Lang15p075102}  & 3.0~\cite{Gesi1975p463}  \\
t-MAPbI$_3$   & 1.70  &2.75 &2.69 &0.60  &1.49   &1.43   &1.45~\cite{Das22p2184}  &1.67~\cite{Umari14p1}    &1.55~\cite{Hao14p8094}      \\
$\gamma$-MAPbI$_3$  &1.76  &2.76 &2.71 &0.72  &1.52  &1.47  &1.74~\cite{Das22p2184}  & -  &1.633~\cite{Ishihara94p269}         \\ 
$\alpha$-MAPbI$_3$  &1.61  &2.66 &2.59 &0.44  &1.34  &1.27  &1.32~\cite{Das22p2184}  &1.675~\cite{Huang16p195211}   &1.55~\cite{Leguy16p6317}   \\
$\alpha$-MAPbBr$_3$ &1.98  &3.27 &3.19 &0.82  &1.97  &1.89  &1.81~\cite{Das22p2184}  & 2.55~\cite{Pandech20p25723}  &2.24~\cite{Leguy16p6317}    \\
$\alpha$-MAPbCl$_3$ &2.48  &3.80 &3.73 &1.28  &2.49  &2.42  &2.41~\cite{Das22p2184}  & 3.49~\cite{Pandech20p25723}  &2.97~\cite{Leguy16p6317}    \\
\bf {MAE} &0.31 &0.98 &0.93 &1.34 &0.24 &0.29 &0.37 &0.24\\
\hline
\bf {MAE$_{t}$}  &0.68  &0.66 &0.63 &1.32  &0.31  &0.37  &0.48    &0.21    \\
\hline
\hline
\end{tabular}
\end{table}

\clearpage
\newpage
\begin{table}[]
\centering
\caption{Comparison of the band gaps (in eV) as computed with different methods and as measured in experiments for $A_2BX_6$ VODPs.}
\label{A2BX6}
\begin{threeparttable}
\begin{tabular}{c|ccc|ccc|cc}
\hline
\hline
\multirow{3}{*}{$A_2BX_6$} & \multicolumn{6}{c|}{This Work} & 
\multicolumn{2}{c}{Reference}  \\
\cline{2-7}\cline{8-9}
 &\multicolumn{3}{c|}{without SOC} &  \multicolumn{3}{c|}{with SOC} & \multirow{2}{*}{HSE(SOC)} & \multirow{2}{*}{Expt} \\
 \cline{2-7}
 & PBE      & ACBN0    &eACBN0 &  PBE      & ACBN0    &eACBN0  \\
\hline
Cs$_2$SnI$_6$        & 0.18  & 0.96 &1.42  &0.13
      &0.80   &1.26    & 0.92~\cite{Dalpian17p025401}    & 1.26~\cite{Lee14p15379}  \\
Cs$_2$SnBr$_6$       & 1.41  & 2.48 &3.18  &1.32      &2.37    &3.07    & 2.49~\cite{Dalpian17p025401}    & 2.7~\cite{Kaltzoglou16p11777}   \\
Cs$_2$SnCl$_6$       & 2.64  & 3.96 &4.61  &2.61      &3.94    &4.59    & 3.89~\cite{Dalpian17p025401}    & 3.9~\cite{Kaltzoglou16p11777}   \\
Cs$_2$TeI$_6$        & 1.39   & 1.82 &2.54  &1.16      &1.58    &2.30    & 1.83~(SOC)~\cite{Maughan16p8453}    & 1.59~\cite{Maughan16p8453}  \\
Cs$_2$TeBr$_6$       & 2.16   & 2.58 &3.34  &2.04      &2.43    &3.19    & 2.7$^a$~(SOC)~\cite{Folgueras21p25126}    & 2.68~\cite{Folgueras21p25126}  \\
Cs$_2$TeCl$_6$       & 2.76   & 3.32 &3.98  &2.61      &3.10    &3.76    & 3.4$^a$~(SOC)~\cite{Folgueras21p25126}    & 3.15~\cite{Folgueras21p25126}  \\
Cs$_2$TiI$_6$        & 0.89  & 1.38  &2.42  &0.76      &1.23    &2.27    & 1.20~($\alpha$ = 0.16)~\cite{Ju18p297}    & 1.02~\cite{Ju18p297}  \\
Cs$_2$TiBr$_6$       & 1.60  & 2.19  &3.45  &1.52      &2.09    &3.35    & 2.01~($\alpha$ = 0.16)~\cite{Ju18p297}    & 1.78~\cite{Ju18p297}  \\
Cs$_2$TiCl$_6$       & 2.27  & 2.93  &4.23   &2.26      &2.90   &4.20    & -       & 2.54~\cite{Kong20p1591} \\
Cs$_2$ZrI$_6$        & 1.83  & 2.40  &3.24  &1.65      &2.21    &3.05    & 2.58~(SOC)~\cite{Cucco21p181903}   & -  \\
Cs$_2$ZrBr$_6$        & 2.83  & 3.58 &4.53  &2.73      &3.48    &4.43    & 3.88~(SOC)~\cite{Cucco21p181903}    & 3.76~\cite{Abfalterer20p1644}  \\
Cs$_2$PdI$_6$        & 0.13  & 0.88  &1.46  &0.11      &0.65    &1.23    & 0.90~\cite{Faizan21p4495}    & 1.41~\cite{Zhou18p2613}  \\
Cs$_2$PdBr$_6$       & 0.76  & 1.72  &2.46  &0.68      &1.60    &2.34    & 1.57~\cite{Wang20p13310}    & 1.6~\cite{Sakai17p6030}  \\
Cs$_2$PdCl$_6$       & 1.42  & 2.50  &3.26  &1.38      &2.46    &3.22    & 2.68~\cite{Wang20p13310}    & 2.4~\cite{Wang20p13310}  \\
Cs$_2$PtI$_6$        & 0.64  & 1.53  &2.19  &0.54      &1.39    &2.05    & 1.39~\cite{Ye22p}    & 1.4~\cite{Schwartz20p2000182}  \\
Cs$_2$PtBr$_6$       & 1.41  & 2.48  &3.30  &1.35      &2.37    &3.19    & 2.33~\cite{Ye22p}    & -  \\
Cs$_2$PtCl$_6$       & 2.01  & 3.11  &3.89  &1.83      &2.91    &3.69    & 3.31~\cite{Ye22p}    & -  \\
\bf {MAE}                  &0.72  &0.24  &0.85   &0.81  &0.22   &0.75    &0.19\\
\hline
\hline
\end{tabular}
\begin{tablenotes}
        \footnotesize
        \item[$a$] Band gap values are extracted from the band structures.
      \end{tablenotes}
  \end{threeparttable}
\end{table}

\clearpage
\newpage
\begin{table}[]
\centering
\caption{Comparison of the band gaps (in eV) as computed with different methods and as measured in experiments for BDHPs. The sc-eACBN0 method simultaneously converge both structure and Hubbard $U$ and $V$ parameters.}
\label{BDHPgaps}
\begin{tabular}{ccccccc}
\hline
\hline
\multirow{2}{*}{} & \multicolumn{4}{c}{This Work} & 
\multicolumn{2}{c}{Reference}  \\
\cline{2-4}\cline{5-7}
& PBE           & ACBN0      &eACBN0     &sc-eACBN0 & HSE     & Expt  \\
\hline
CsTlCl$_3$      & 1.00    & 1.50          & 1.89    & 2.67 & 1.3~\cite{Retuerto13p4071}    &2.5~\cite{Retuerto13p4071} \\
CsTlF$_3$       & 2.15    & 2.64          & 3.10    & 4.46 & 3.9~\cite{Retuerto13p4071}    & - \\
CsAuCl$_3$      & 0.95    & 0.90          & 1.37    & 1.73 & 1.51~\cite{Dalpian18p075135}   & 2.04~\cite{Liu99p7925}    \\
\hline
\hline
\end{tabular}
\end{table}

\begin{table}[]
\centering
\caption{Self-consistent $U$ and $V$ values (in eV) for different orbitals in CsAuCl$_3$ comptuted with ACBN0, eACBN0, and sc-eACBN0. $V_{dp}$ is the intersite Hubbard term between nearest-neighboring Au-$5d$ and Cl-$3p$ orbitals. Underlined Hubbard parameters are most important for band gap predictions.}
\label{UVvales_BDHP}
\begin{tabular}{ccccc}
\hline
\hline
\multirow{2}{*}{$U/V$ values} & \multicolumn{3}{c}{CsAuCl$_3$} \\
\cline{2-4}
& ACBN0      &eACBN0     & sc-eACBN0  \\
\hline
$U$(Cs-6$s$)       & 0.00    & 0.00     & 0.00        \\
$U$(Cs-5$p$)       & 2.87    & 2.85     & 2.39       \\
$U$(Au$^{\rm L}$-6$s$) & 0.11     & 0.15     & 0.22 \\
\underline{$U$(Au$^{\rm L}$-5$d$)} & 2.23    & 2.35     & 2.59\\
$U$(Au$^{\rm S}$-6$s$) & 0.05     & 0.06     & 0.06 \\
\underline{$U$(Au$^{\rm S}$-5$d$)} & 3.69    & 3.49     & 3.53\\
\underline{$U$(Cl$^{\rm a}$-3$p$)} & 3.42    & 3.44     & 3.52        \\
\underline{$U$(Cl$^{\rm e}$-3$p$)} & 3.60    & 3.65     & 3.79     \\
\underline{$V_{dp}$(Au$^{\rm L}$-Cl$^{\rm a}$)} & -       & 2.56     & 2.59    \\
\underline{$V_{dp}$(Au$^{\rm L}$-Cl$^{\rm e}$)} & -       & 1.96     & 1.88      \\
\underline{$V_{dp}$(Au$^{\rm S}$-Cl$^{\rm a}$)} & -       & 1.68     & 1.67      \\
\underline{$V_{dp}$(Au$^{\rm S}$-Cl$^{\rm e}$)} & -       & 2.32     & 2.35      \\
\hline
\hline
\end{tabular}
\end{table}
\begin{table}[]
\centering
\caption{Theoretical and experimental lattice constants and Au-Cl bond lengths (in \AA) of CsAuCl$_3$.}
\label{CsAuCl3_bonds}
\begin{tabular}{ccccc}
\hline
\hline
 &PBE & sc-eACBN0 &Previous Work~\cite{Winkler01p214103} &Expt.~\cite{Denner79p360}\\
\hline
$a$ & 7.62  &7.82 & 7.6037 &7.495\\ 
$c$   & 11.31 &11.39 & 11.251 &10.880\\
\hline
Au$^{\rm L}$-Cl$^{\rm a}$ &2.31 &2.31 &2.314 &2.28 \\
Au$^{\rm L}$-Cl$^{\rm e}$ &3.04 &3.19 &3.018 &3.01 \\
Au$^{\rm S}$-Cl$^{\rm a}$ &3.34 &3.38 &3.311 &3.15 \\
Au$^{\rm S}$-Cl$^{\rm e}$ &2.35 &2.33 &2.359 &2.29 \\
\hline
\hline
\end{tabular}
\end{table}

\clearpage
\newpage

\clearpage
\newpage

\clearpage
\newpage

\clearpage
\newpage

\clearpage
\newpage

\clearpage
\newpage

\clearpage
\newpage

\clearpage
\newpage

\clearpage
\newpage

\clearpage
\newpage

\clearpage
\newpage

\clearpage
\newpage

\end{document}